**Tipografía Time New Roman 12 ptos.        Interlineado a 1.5     Justificación completa para el cuerpo**

# Herramientas tecnológicas en Android para la formación de mapeadores y promotores de Mapa Verde

## Technological tools in Android for the training of mappers and promoters of Green Map

Ing. Yosvany Medina Carbó [1*] https://orcid.org/0000-0003-3590-0706

Dr.C. Álvaro Celestino Alonso Vázquez [1] https://orcid.org/0000-0002-9895-5790

MSc. Reina María Rodríguez García[1] https://orcid.org/0000-0003-4533-279X

[1] Centro Universitario Municipal ¨Hermanos Saíz Montes de Oca¨ Consolación del Sur. Calle 53 No 6808 e/ 68 y 70 Consolación del Sur, Pinar del Río, Cuba.

*Autor para la correspondencia: yosvany.medina@upr.edu.cu

**RESUMEN**

Cuando se habla de tecnologías y medio ambiente, normalmente, se imagina gran cantidad de equipos, técnicas, tecnologías y herramientas contaminando el entorno natural. Se lleva años proyectando sobre el planeta las consecuencias buenas y malas de nuestro desarrollo, y parte de ese desarrollo se ve reflejado en las nuevas tecnologías, entre las cuales se encuentra el teléfono móvil. En el municipio de Consolación del Sur y desde el Centro Universitario Municipal se trabaja en el proyecto "Implementación de la Metodología de Mapa Verde en la gestión de la educación ambiental en comunidades consolareñas" para la formación de una cultura ambiental para el desarrollo sostenible creando conciencia del cuidado y protección del entorno.

**Tipografía Time New Roman 12 ptos.     Interlineado a 1.5     Justificación completa para el cuerpo**

El presente trabajo está dado a solucionar el siguiente problema: ¿cómo contribuir en la construcción de un paquete de herramientas informáticas para la implementación de la metodología de Mapa Verde en la gestión ambiental en comunidades consolareñas y la formación de mapeadores y promotores de Mapa Verde para el desarrollo de los mapas verdes de las comunidades del municipio Consolación del Sur? Para ello se desarrolló dos aplicaciones en Android para dispositivos móviles basadas en la metodología de Mapa Verde y dar respuesta así al siguiente objetivo: Desarrollar un paquete de aplicaciones informáticas para la implementación de la metodología de Mapa Verde en la gestión de la educación ambiental en comunidades consolareñas y la formación de mapeadores y promotores de Mapa Verde que permita el desarrollo de los mapas verdes de las comunidades del municipio Consolación del Sur.
**Palabras clave**: aplicación informática; dispositivos móviles; Mapa Verde; medio ambiente; metodología.

 ABSTRACT

When you talk about technologies and the environment, you usually imagine a lot of equipment, techniques, technologies and tools polluting the natural environment. The good and bad consequences of our development have been projected on the planet for years, and part of that development is reflected in the new technologies, among which is the mobile phone. In the municipality of Consolación del Sur and from the Municipal University Center, the project "Implementation of the Green Map Methodology in the management of environmental education in console communities" for the formation of an environmental culture for sustainable development is created, creating awareness of care and protection of the environment. The present work is given to solve the following problem: how to contribute in the construction of a package of computer tools for the implementation of the Green Map methodology in environmental management in console communities and the training of mappers and promoters of Green Map for the development of green maps of the communities of the municipality Consolación del Sur? For this purpose, two Android applications for mobile devices based on the Green Map methodology were developed, thus responding to the following objective: Develop a package of computer applications for the implementation of the Green Map methodology in the management of environmental education in console communities and the formation of mappers and promoters of the Green Map that allows the development of the green maps of the communities of the Consolación del Sur municipality.

# Introducción

**Tipografía Time New Roman 12 ptos.        Interlineado a 1.5     Justificación completa para el cuerpo**

La introducción y el uso de las Nuevas Tecnologías de la Información y las Comunicaciones (NTICs) ha significado a escala mundial un salto vertiginoso en el desarrollo científico técnico; desde su llegada a los escenarios nacionales, se han convertido en un elemento indispensable para establecer las líneas de desarrollo de la sociedad cubana, buscando dar solución a los problemas del hombre. Tomando en consideración esa realidad mundial y gracias a la firme voluntad política del gobierno cubano, se han desarrollado en el país múltiples programas encaminados a lograr la informatización de la sociedad; que están relacionados con el proceso de utilización ordenada y masiva de las TIC para satisfacer las necesidades de información y conocimiento de la sociedad lo que constituye una de las metas que tiene Cuba en el presente y para los próximos años.

Una de las líneas de desarrollo de la sociedad cubana está relacionada con el cuidado y protección del medio ambiente, así como la Tarea Vida, programa que tiene Cuba para enfrentar el cambio climático, la cual debe estar estrechamente relacionada con el desarrollo de las TIC y el proceso de informatización de la sociedad cubana creando para ello aplicaciones informáticas encaminadas a fomentar una cultura ambiental para el desarrollo sostenible. Es por ello que cuidar el medio ambiente es un reto enorme para la sociedad cubana del siglo XXI, constituyendo una tarea de todos, y cada vez más la sociedad está concienciada de su importancia, como legado para las generaciones futuras. Como decía aquel eslogan que se hizo tan popular hace unos años, el reto consiste en actuar localmente para cambiar globalmente. Si todos nos concienciamos de la importancia del medio ambiente, un futuro verde se proyectará sobre las generaciones de los años venideros. La naturaleza debe ser cuidada debidamente; el cambio climático es un problema que afecta a sociedades de todo el mundo y es nuestro deber reducir sus efectos mediante una vida respetuosa con el medio ambiente.

Las TIC pueden ayudar a facilitar y dinamizar el aprendizaje, y un ejemplo son las aplicaciones para dispositivos móviles que pueden resultar muy útiles para llevar a cabo educación ambiental en las aulas sin que se convierta en una tarea aburrida y compleja. Las aplicaciones móviles sirven para fines específicos dentro de una gran variedad de áreas, para fines personales, comerciales, industriales, etc; su utilización está

**Tipografía Time New Roman 12 ptos.        Interlineado a 1.5    Justificación completa para el cuerpo**

presente en diferentes contextos, desde la salud (Aanensen, y otros, 2009) (Hui, y otros, 2017) (Cruz-Barragán, y otros, 2017) hasta la Bibliotecología, la Ciencia de la Información (Arroyo-Vásquez, 2011) y el medio ambiente. Hoy, gracias a los dispositivos móviles y las aplicaciones verdes, el desafío parece un poco más sencillo.

El uso de herramientas móviles, conocidas también como aplicaciones informáticas ejecutadas desde teléfonos celulares (App), es una realidad latente e ineludible que forma parte de las dinámicas y procesos cotidianos de la vida; además de estar presente en la educación e investigación y particularmente en la gestión de información de toda índole (personal, académica o laboral o de entretenimiento). Dicho fenómeno suscita especial interés respecto a la necesidad y pertinencia de gestionar la información en tiempo real, como respuesta al creciente interés por la inmediatez en el acceso a la información desde cualquier punto. (Cebrián, 2009)

Con el objetivo de fomentar la Educación Ambiental en las comunidades del municipio de Consolación del Sur se comenzó a trabajar desde el Centro Universitario Municipal "Hermanos Saíz Montes de Oca" en el proyecto Programa Interdisciplinario de Cultura Ambiental (PICA), el cual tenía como objetivo elevar la cultura ambiental de las comunidades consolareñas a partir de la educación, para la promoción de un desarrollo local sostenible. Este proyecto contribuyó a la formación de aprendizajes desarrolladores en educadores ambientales y actores sociales sobre los temas: Mapa Verde, Economía Solidaria, Responsabilidad Social Empresarial, reciclaje, género, violencia, alcoholismo, cambio climático, vulnerabilidades y riesgos, enfermedades y droga, que conduzcan a estilos de vidas en armonía con el medio ambiente. Con el progresivo deterioro del planeta cada vez es más importante la tarea de los docentes a la hora de enseñar a los pequeños y jóvenes a ser respetuosos con el medio ambiente.

Para ello el Centro Universitario Municipal "Hermanos Saíz Montes de Oca" del municipio Consolación del Sur con la colaboración del Centro Félix Varela y la Red Nacional de Mapa Verde comenzó a trabajar con la metodología de Mapa Verde, para la formación de educadores ambientales, mapeadores y promotores de Mapa Verde en las comunidades educando y creando conciencia del cuidado y conservación del entorno.

**Tipografía Time New Roman 12 ptos.         Interlineado a 1.5     Justificación completa para el cuerpo**

Hoy en día el Centro Universitario Municipal lugar donde se encuentra el nodo de Mapa Verde en el municipio cuenta con un proyecto denominado "Implementación de la Metodología de Mapa Verde en la gestión de la educación ambiental en comunidades consolareñas", el cual contribuye a la formación de aprendizajes desarrolladores en educadores ambientales y actores sociales sobre los temas: Mapa Verde, Economía Solidaria, Responsabilidad Social Empresarial, reciclaje, género, violencia, alcoholismo, cambio climático, vulnerabilidades y riesgos, enfermedades y droga, que conduzcan a estilos de vidas en armonía con el medio ambiente.

Cuando se va a desarrollar el mapa verde de cada comunidad es necesario que cada uno de los mapeadores realice un diagnóstico de las condiciones ambientales de su comunidad y conocer cada uno de los íconos que pueden estar presentes en la misma, así como donde utilizarlos y el significado de cada uno de ellos. El nodo apoyado en los materiales que forman parte de la caja de herramientas de los promotores(as) cubanos(as), los cuales constituyen fuente obligada de consulta trabaja en la capacitación de los actores sociales, decisores, delegados de circunscripción, presidentes de consejo popular, líderes comunitarios, niños, niñas, jóvenes, hombres, mujeres, ancianos y ancianas de las comunidades en la formación de los mapeadores con el conocimiento y estudio de las condiciones ambientales detectadas en cada comunidad y el plegable con los íconos y su significado utilizando para ello los libros entregados por el centro Félix Varela, los talleres y la capacitación recibida por la red nacional de Mapa Verde, así como toda la información relacionada con la metodología y con la iconografía utilizada para la confección de los mapas.

Todos estos materiales con que cuenta el Centro Universitario Municipal para el estudio de la iconografía de la metodología Mapa Verde como son libros, plegables, boletines, etc. en algunas ocasiones no alcanzan para que los mapeadores puedan trabajar en su comunidad todo el tiempo. Esto último trae consigo que en varias ocasiones los mapeadores no estén seguros del ícono que van a utilizar en cada sitio verde detectado por no contar con los materiales necesarios para trabajar, lo que influye en la desmotivación de muchos de ellos en querer transformar la comunidad en que viven y dificulta la realización de las acciones a tomar en cuenta, por parte de los decisores.

**Tipografía Time New Roman 12 ptos.**     **Interlineado a 1.5**     **Justificación completa para el cuerpo**

Producto a esta situación todos los materiales con que cuenta el nodo para el estudio de la iconografía de la metodología Mapa Verde como son libros, plegables en algunas ocasiones no alcanzan para que los mapeadores puedan trabajar en su comunidad todo el tiempo. Esto último trae consigo que en varias ocasiones los mapeadores no estén seguros del ícono que van a utilizar en cada sitio verde detectado por no contar con los materiales necesarios para trabajar, lo que influye en la desmotivación de muchos de ellos en querer transformar la comunidad en que viven y dificulta la realización de las acciones a tomar en cuenta, por parte de los decisores.

A partir de la problemática planteada se define como problema a resolver: ¿cómo contribuir en la construcción de un paquete de herramientas informáticas para la implementación de la metodología de Mapa Verde en la gestión ambiental en comunidades consolareñas y la formación de mapeadores y promotores de Mapa Verde para el desarrollo de los mapas verdes de las comunidades del municipio Consolación del Sur?

Para dar solución al problema planteado se trazó como Objetivo general: Desarrollar un paquete de aplicaciones informáticas para la implementación de la metodología de Mapa Verde en la gestión de la educación ambiental en comunidades consolareñas y la formación de mapeadores y promotores de Mapa Verde que permita el desarrollo de los mapas verdes de las comunidades del municipio Consolación del Sur.

## Métodos o Metodología Computacional

Los principales métodos empleados en la investigación son los siguientes:

**El método dialéctico-materialista** como base metodológica que definirá el uso de los métodos científicos generales, plantea como principio: el movimiento espiral y ascendente del conocimiento, identificando las contradicciones, los nexos y las transformaciones que se evidencian a lo largo de la investigación. A través

**Tipografía Time New Roman 12 ptos.        Interlineado a 1.5        Justificación completa para el cuerpo**

del mismo se podrá determinar las particularidades, regularidades y tendencias del proceso de educación ambiental en la realidad internacional, cubana y el empleo de herramientas como la Metodología Mapa Verde para elevar la cultura ambiental comunitaria en el municipio Consolación del Sur, en particular, para establecer la consecuente periodización histórica. Este método además permitirá establecer el marco conceptual sobre la educación ambiental y la educación ambiental comunitaria, y las aplicaciones para móviles con sistemas operativos Android (versión 4.2 o superior) lo que favorecerá la integración de los elementos teóricos que sustentan la concepción sobre este proceso.

Se utilizarán métodos teóricos como:

El **método de análisis histórico y lógico**. Este permitirá el estudio de las distintas etapas por las que transitara el debate ambientalista y el desarrollo de las TIC antes de desembocar en el actual proceso de internacionalización, así como la trayectoria concreta que ha seguido la informatización de la sociedad cubana en el particular de la educación ambiental y la gestión ambiental para el desarrollo sostenible. Permitirá establecer el marco conceptual de la investigación, así como determinar las principales tendencias del proceso de educación ambiental comunitaria con auxilio de las TIC.

Como **métodos empíricos** esenciales se utilizará **La medición**: como método que se utiliza para obtener información numérica acerca de la cualidad de los mapeadores, promotores y líderes comunitarios en su accionar por la educación ambiental, donde se comparan magnitudes medibles y conocidas. Es la atribución de valores cuantitativos a determinadas propiedades relacionadas con el proceso de las TIC en la informatización de la población cubana en función de la gestión de la educación ambiental comunitaria. Su uso suele aparecer combinado con la técnica de las entrevistas y las encuestas

**Como técnicas de investigación se emplearon: la** entrevista, la encuesta **y el análisis documental** que permitirán el diagnóstico del objeto de estudio, la demostración del problema, así como propuestas de solución al mismo.

**Tipografía Time New Roman 12 ptos.     Interlineado a 1.5     Justificación completa para el cuerpo**

las entrevistas y las encuestas: Se utilizan para identificar los problemas ambientales, para la obtención de información mediante el diálogo en el que se transforma y sistematiza la información conocida por estas Se utilizará además las entrevistas a mapeadores, promotores y líderes comunitarios.

**Método de investigación acción participativa (IAP)**, permitirá promover la participación activa de los sujetos implicados en la transformación y el mejoramiento de los estilos de vida pertenecientes al Nodo Mapa Verde de Consolación del Sur, a partir de la potenciación de las aplicaciones móviles para sistemas operativos Android que le dará un carácter educativo de la investigación.

Se emplearán como procedimientos: el análisis y síntesis y la inducción-deducción para la determinación de los fundamentos teóricos- metodológicos del proceso de educación ambiental.

El análisis permitirá el estudio del comportamiento del trabajo de los mapeadores en su accionar para contribuir a la educación ambiental, el compromiso e integración en la comunidad con apoyo y utilización de las TIC. Mediante la síntesis se descubrirá la interacción y el condicionamiento mutuo, que ejercen en el modo de actuación, los miembros del nodo Mapa Verde en el proceso de educación ambiental.

En la investigación se utilizarán herramientas de programación para la confección de las aplicaciones y estadísticas para el procesamiento de datos, en particular, los procedimientos de la estadística descriptiva y en la elaboración de gráficos.

## Metodología de desarrollo de software utilizada para el desarrollo del paquete de herramientas.

La metodología elegida para guiar el desarrollo del paquete de herramientas sobre la metodología de Mapa Verde es Mobile-D. El objetivo de este método es conseguir ciclos de desarrollo muy rápidos en equipos muy pequeños (de no más de diez desarrolladores) trabajando en un mismo espacio físico. Fue creado en un proyecto finlandés por investigadores de la VIT (Instituto de Investigación Finlandés) en 2005, pero sigue

**Tipografía Time New Roman 12 ptos.        Interlineado a 1.5     Justificación completa para el cuerpo**

estando vigente. Basado en metodologías conocidas pero aplicadas de forma estricta como: extreme programming (XP), Crystal Methodologies y Rational Unified Process (RUP), XP para las prácticas de desarrollo, Crystal para escalar los métodos y RUP como base en el diseño del ciclo de vida. En esta metodología el ciclo del proyecto se divide en cinco fases: exploración, inicialización, productización o fase de producto, estabilización y prueba del sistema [Gómez, y otros, 2016].

### Fases de la Metodología Mobile-D

El ciclo del proyecto se divide en cinco fases:

- Exploración: Se dedica a la planificación y a los conceptos básicos del proyecto. Es diferente del resto de fases.
- Inicialización: Se preparan e identifican todos los recursos necesarios. Se establece el entorno técnico.
- Productización o fase de producto: se repiten iterativamente las subfases, con un día de planificación, uno de trabajo y uno de entrega.
- Fase de estabilización: Se llevan a cabo las acciones de integración para asegurar que el sistema completo funciona correctamente.
- Fase de pruebas y reparación: Tiene como meta la disponibilidad de una versión estable y plenamente funcional del sistema según los requisitos del cliente [Ramírez, 2016].

## Herramientas y entornos de trabajo utilizados para el desarrollo del paquete de herramientas

**Android Studio** es un IDE (por sus siglas en inglés, Entorno de Desarrollo Integrado) oficial para la plataforma Android totalmente gratuito a través de la Licencia Apache 2.0 y multiplataforma (Microsoft Windows, macOS y GNU/Linux). Android Studio está basado en el software IntelliJ IDEA de JetBrains y fue lanzado como reemplazo a Eclipse como el IDE oficial para el desarrollo de aplicaciones para Android. Esta aplicación ofrece una herramienta completa para desarrollar y depurar aplicaciones para el sistema

**Tipografía Time New Roman 12 ptos.     Interlineado a 1.5     Justificación completa para el cuerpo**

operativo de Google para dispositivos móviles. Con ella podremos realizar la edición de código, depuración, utilizar herramientas de rendimiento, cuenta con un sistema flexible de compilación y creación y despliegue instantáneo, que le permite centrarse en la creación de aplicaciones. (Fuentes, y otros, 2016)

**BlueStacks App Player.** Es una herramienta que permite ejecutar aplicaciones Android en Windows y Macintosh capaz de correr miles de aplicaciones móviles en la comodidad de la pantalla grande a través de una cómoda interfaz. Fue desarrollada por la compañía de tecnología estadounidense BlueStack, que fue fundada en 2009 por Rosen Sharma, ex director de tecnología en McAfee y miembro de Cloud.com. La aplicación cuenta entre sus inversores a Samsung, Intel y otras grandes compañías de tecnología. Su historia se remonta al año 2011, cuando decidieron extender los límites de las aplicaciones para móviles y crear una herramienta que permitiera ejecutar aplicaciones Android en sistemas operativos Windows o Mac.

**Lenguaje de programación Java.** Java es un lenguaje de programación de propósito general, concurrente, orientado a objetos que fue diseñado específicamente para tener tan pocas dependencias de implementación como fuera posible. Su intención es permitir que los desarrolladores de aplicaciones escriban el programa una vez y lo ejecuten en cualquier dispositivo (conocido en inglés como WORA, o "write once, run anywhere"), lo que quiere decir que el código que es ejecutado en una plataforma no tiene que ser recompilado para correr en otra. Java es, a partir de 2012, uno de los lenguajes de programación más populares en uso, particularmente para aplicaciones de cliente-servidor de web, con unos 10 millones de usuarios reportados. Es un lenguaje de programación orientado a objetos muy universal que se ejecuta en la máquina virtual de Java.  Fue originalmente desarrollado por James Gosling, de Sun Microsystems (constituida en 1983 y posteriormente adquirida el 27 de enero de 2010 por la compañía Oracle), y publicado en 1995 como un componente fundamental de la plataforma Java de Sun Microsystems. Java es rápido, seguro y fiable.

### Patrón arquitectónico para realizar el paquete de herramientas.

**Tipografía Time New Roman 12 ptos.          Interlineado a 1.5     Justificación completa para el cuerpo**

Un patrón arquitectónico define la estructura básica de una aplicación, provee un subconjunto de subsistemas predefinidos, incluyendo reglas, lineamientos para conectarlos y pautas para su organización, además constituye una plantilla de construcción.

- Entre las ventajas del uso de patrones, se pueden encontrar:
- Permiten la reutilización de soluciones arquitectónicas de calidad.
- Son de gran ayuda para controlar la complejidad de un diseño.
- Facilitan la documentación de diseños arquitectónicos.
- Proporcionan un vocabulario común que mejora la comunicación entre diseñadores.

Los patrones arquitectónicos expresan una organización estructural para un sistema, permiten estructurar los componentes y subsistemas de un sistema. La abstracción más alta en cuanto a soluciones a través de patrones se obtiene a través del uso de patrones de arquitectura.

### Patrón arquitectónico N-Capas.

Se propone el uso del patrón arquitectónico N-Capas para el desarrollo de la aplicación, ya que el entorno de ejecución estará distribuido en diferentes elementos o nodos, tanto lógicos como físicos. Aquí aparecen dos conceptos importantes: las capas que se encargan de la distribución lógica de los componentes, y los niveles que se refieren a la colocación física de los recursos. La característica principal de este patrón es que cada capa oculta las capas inferiores de las siguientes superiores a esta. (Muñoz, 2018)

La descomposición en capas que se propone es:
- Capa de presentación: es la única que interactúa directamente con el usuario presentándole el sistema, permitiendo el intercambio de información entre ambos. Contiene una interfaz gráfica que posibilita mostrar a los usuarios los resultados de sus peticiones y estados de la aplicación. Esta capa solo interactúa con la capa de negocio, que es su inferior inmediata.
- Lógica de programación: es la que recibe las peticiones de la capa de presentación y le envía las respuestas tras el proceso. Aquí se realiza la mayor parte del procesamiento de la información del

**Tipografía Time New Roman 12 ptos.         Interlineado a 1.5     Justificación completa para el cuerpo**

dominio de la aplicación, donde se ejecutan los algoritmos específicos del programa. Esta capa interactúa hacia arriba, con la capa de presentación, para recibir sus solicitudes y presentarle los resultados al usuario e interactúa hacia abajo haciendo solicitudes a la capa de red y a la capa de acceso de datos.

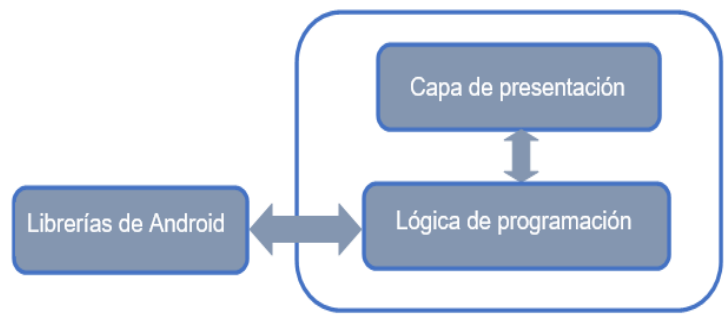

**Fig. 1.** Diagrama de la arquitectura del proyecto.

A continuación, se describen los elementos de cada capa:

- Capa de presentación: En esta capa se encuentra la actividad principal de la aplicación que contiene los componentes visuales que se muestran al usuario.
- Lógica de programación: En esta capa se representan todos los componentes lógicos del flujo de trabajo, o sea las clases Javas que se encargan de ejecutar las tareas propias de la aplicación.

# Resultados y discusión

El Mapa Verde es la representación del ambiente natural y cultural que caracteriza el entorno, constituye una metodología de diagnóstico, planificación y gestión ambiental participativa, promueve una efectiva participación en la búsqueda de alternativas de solución a problemáticas locales de una manera creativa, que influye positivamente en las esferas educativa y ambiental. La metodología de Mapa Verde es eficaz para fomentar programas de Educación Ambiental, encaminados al desarrollo de comunidades sostenibles que

**Tipografía Time New Roman 12 ptos.        Interlineado a 1.5      Justificación completa para el cuerpo**

generan acciones mediante un proceso educativo, participativo y creativo; es un medio de consulta, información y actuación donde participan: niños, jóvenes, adultos, personas de la tercera edad, profesionales, todas las organizaciones interesadas, comunidades rurales y urbanas, otros proyectos de Educación Ambiental, decidores, en fin, todo aquel que desee vivir en un mundo mejor, guiados por el lema "Solo se aprende a mapear, mapeando".(Bidart, y otros, 2017)

El Mapa Verde llega a Cuba en junio de 1998 con la Conferencia internacional "Ética y Cultura del Desarrollo Sostenible: Construyendo una Economía Sostenible" convocada por la Oficina Regional de la UNESCO, el American Friends Service Committee (USA) y el Centro Félix Varela (Cuba), en la cual la señora Wendy Brawer (creadora de esta metodología) presentó una ponencia sobre el Mapa Verde como promotor del ordenamiento ecológico en los habitantes urbanos, y como un estímulo para el descubrimiento por parte de niños, niñas, jóvenes y adultos de los eco recursos en sus comunidades. En la comisión donde se presentó esta ponencia participaron integrantes del Centro Félix Varela (CFV) los cuales acordaron, con la Junta directiva, que esta era una valiosa herramienta de educación ambiental y de conciencia ciudadana, aplicable a nuestro contexto. (Bidart, y otros, 2017)

La práctica de la implementación del Mapa Verde en Cuba ha generado la producción de materiales comunicativos, entre los que se encuentran: Carta al Mapeador (boletín de la Red surgido en 2002 como respuesta a la solicitud de los mapeadores, cuatro números); los documentales A mapear un Sueño (2002), que muestra, a través de la experiencia de niños y niñas mapeadores el proceso de elaboración del Mapa Verde, por tal razón se convirtió en una herramienta eficaz para el trabajo de la Red, y Gotica a Gotica (2004), a través del cual se refleja el proceso de gestión local en algunas experiencias comunitarias a partir del trabajo con el Mapa Verde; manual de trabajo Mapa Verde: una mirada al desarrollo local; Mapeando nuestra tierra común, una guía para el mapeo verde y comunitario, resultado de la colaboración con la Red Panamericana que ofrece herramientas para el mapeo y resultados de experiencias nacionales y foráneas; Caminando por el Mapa Verde de Cuba, publicación de la sistematización del proceso de capacitación en la Red de Mapa Verde; Repensar un Sueño, publicación de la sistematización del proceso de participación en

**Tipografía Time New Roman 12 ptos.       Interlineado a 1.5     Justificación completa para el cuerpo**

la Red de Mapa Verde. Estos materiales forman parte de la caja de herramientas de los promotores(as) cubanos(as) y constituyen fuente obligada de consulta para el trabajo, permiten legitimar el Mapa Verde en nuestro país y confrontarlo con experiencias externas.

Para confeccionar el mapa verde de una comunidad se hace necesario conocer una serie de términos importantes como: (Ventosa, y otros, 2015)

- ✓ **Sitios Verdes**: Son todos aquellos lugares que tienen trascendencia e importancia social para los habitantes de la comunidad.
- ✓ **Mapeo comunitario**: Confección de un mapa desde la comunidad. Es un proceso creativo y divertido a través del cual se demuestra cómo perciben los habitantes de la comunidad, el lugar donde viven los que participan, se inicia con la formación de grupos mapeadores que recorren los sitios de interés que deciden llevar al mapa, realizando una especie de levantamiento con su nombre, dirección e icono correspondiente.
- ✓ **Íconos**: Sistema de símbolos internacionales que permite reconocer e identificar los considerados como sitios verdes en la comunidad.

Los íconos de Mapa Verde se utilizan para identificar y caracterizar lugares significativos de la comunidad relacionados con la sociedad y el medio natural que marcan una vía o tendencia hacia la sustentabilidad (sitios verdes). Además, están estructurados por los siguientes géneros: Vida Sustentable, Naturaleza y Cultura y Sociedad, y estos géneros a su vez en categorías encontrándose dentro de Vida Sustentable (Economía Verde, Movilidad, Peligros y desafíos y Tecnología y diseño), en Naturaleza encontramos a las categorías (Tierra y agua, Flora, Fauna, Actividades al aire libre) y en el género Cultura y Sociedad tenemos a las categorías (Característica Cultural, Eco información, Justicia y activismo, Obras públicas y referencia). Para dar cumplimiento al objetivo del presente trabajo se presenta la aplicación informática para dispositivos móviles Metodología de Mapa Verde en Cuba (Figura 2), la cual está encaminada a elaborar el Mapa Verde de cada una de las comunidades del municipio Consolación del Sur y transmitir una adecuada

**Tipografía Time New Roman 12 ptos.        Interlineado a 1.5      Justificación completa para el cuerpo**

educación ambiental a los moradores que conviven en ellas y la aplicación informática Memorizador de Íconos de Mapa Verde (Figura 7) consistente en un juego de memoria con los íconos de la metodología.

La aplicación Metodología de Mapa Verde en Cuba constituye la versión 1.0 de una aplicación para dispositivos móviles desarrollada en Android sobre dicha metodología, la cual tiene como propósito explicar aspectos relacionados con la misma tales como: ¿Qué es el Mapa Verde?, ¿Cómo se hace el Mapa Verde?, Iconografía de Mapa Verde, Mapa Verde en Cuba, Red de Mapa Verde en Cuba, Elementos a considerar en el diseño de mapas y logos, así como otros elementos de interés presentes en la metodología. Para ello la aplicación está compuesta por un menú principal donde al interactuar con él el usuario navega por cada uno de los temas relacionados anteriormente, además de contar con un glosario de términos usados muy frecuentemente en nuestro trabajo como promotores y multiplicadores de Mapa Verde. Esta aplicación puede ser instalada en todos los teléfonos celulares con sistema operativo Android en su versión 4.2 o superior y en todas las escuelas donde existan mapeadores con la ayuda de un emulador de Android para PC y en coordinación con los técnicos de computación ejercitar a los niños en el conocimiento de la metodología.

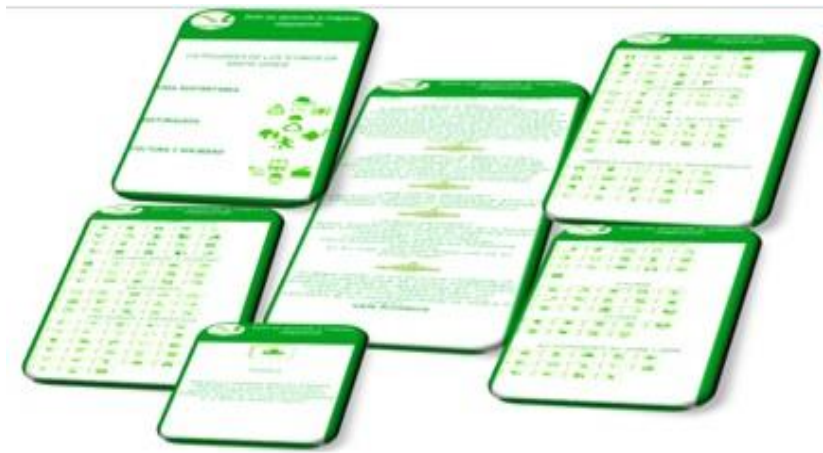

**Fig. 2.** Aplicación donde se explica la metodología Mapa Verde. (Fuente: Elaboración propia)

**Tipografía Time New Roman 12 ptos.          Interlineado a 1.5      Justificación completa para el cuerpo**

En la figura 3 se muestran algunas capturas de pantalla de la aplicación relacionadas con la historia de la metodología de Mapa Verde, donde se puede ver ¿Qué es el Mapa Verde?, el Mapa Verde en Cuba y la Red de Mapa Verde en Cuba.

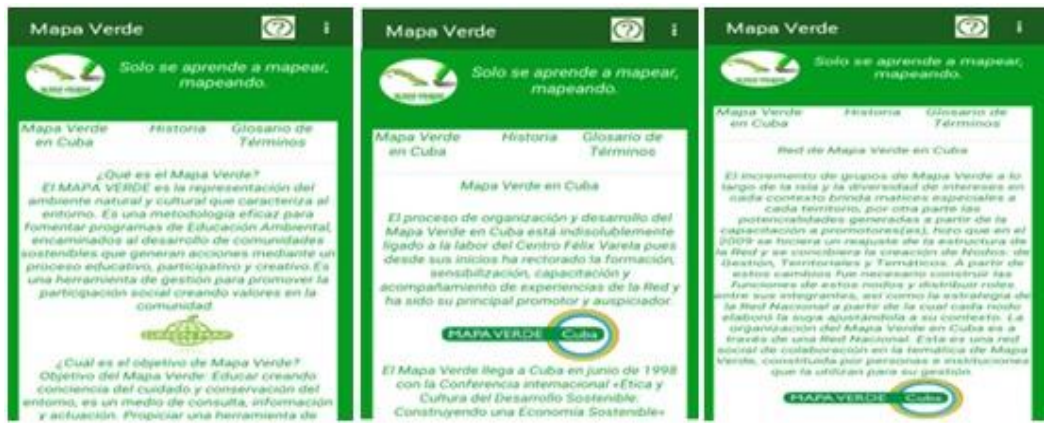

**Fig. 3.** Captura de pantalla de la aplicación con la metodología de Mapa Verde. (Fuente: Elaboración propia)

En la figura 4 se muestran las capturas de pantalla de la aplicación relacionadas con los elementos a considerar en el diseño de mapas y logos, así como la construcción del mapa paso a paso y la iconografía utilizada por la metodología de Mapa Verde para el desarrollo de cada uno de los mapas.

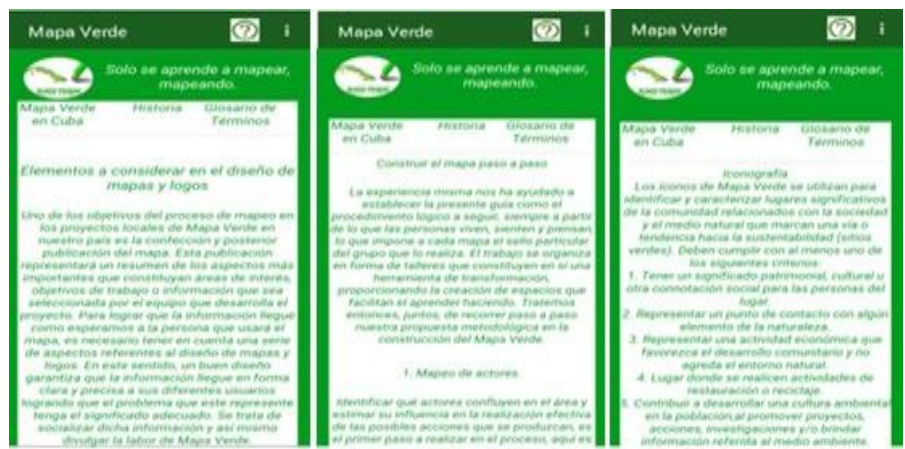

**Fig. 4.** Captura de pantalla de la aplicación con la metodología de Mapa Verde. (Fuente: Elaboración propia)

**Tipografía Time New Roman 12 ptos.　　　Interlineado a 1.5　　　Justificación completa para el cuerpo**

En la figura 5 se muestran las capturas de pantalla de la aplicación relacionadas con los géneros de los íconos de la metodología de Mapa Verde, así como la explicación de cada uno de estos géneros y además las categorías presentes en cada uno de ellos.

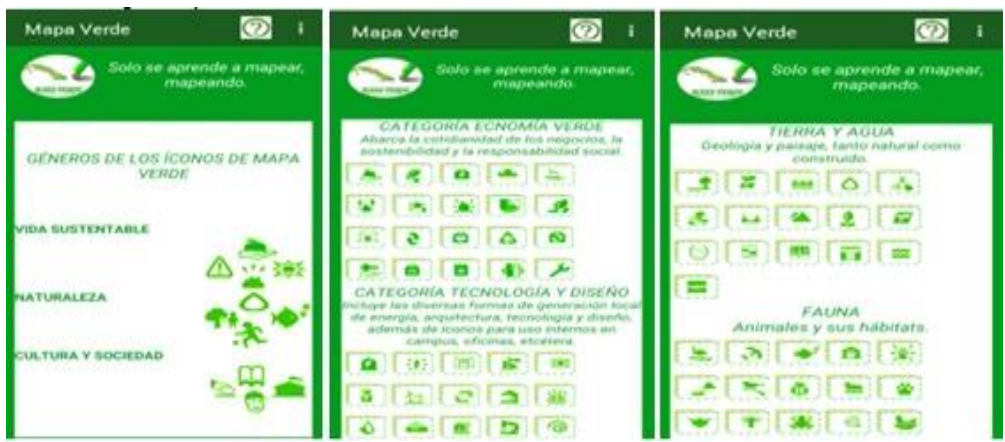

**Fig. 5.** Captura de pantalla de la aplicación con la metodología de Mapa Verde. (Fuente: Elaboración propia)

En la figura 6 se muestran las capturas de pantalla de la aplicación relacionadas con la explicación de cada uno de los géneros de íconos de la metodología de Mapa Verde y además las categorías presentes en cada uno de ellos, así como la explicación de cada uno de los íconos y el glosario de términos utilizado por la metodología.

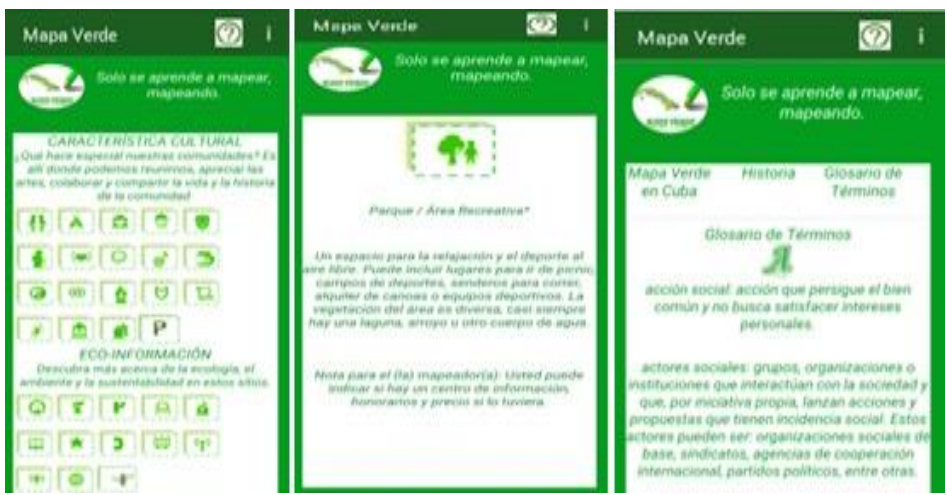

**Fig. 6.** Captura de pantalla de la aplicación con la metodología de Mapa Verde. (Fuente: Elaboración propia)

**Tipografía Time New Roman 12 ptos.     Interlineado a 1.5     Justificación completa para el cuerpo**

La aplicación desarrollada en Android y que lleva por nombre Memorizador de Íconos de Mapa Verde es la versión 1.0 de un juego de memoria con los Íconos de la metodología Mapa Verde que le permite comprobar cómo está su memoria, así como ayuda a poner en práctica su nivel de concentración; consistente en encontrar los pares de íconos iguales. El juego está dividido en cada uno de los géneros de íconos de la metodología y estos a su vez en las categorías de cada uno. Es un excelente ejercicio para el entrenamiento de la memoria de las personas que puede instalarse en todos los teléfonos celulares con sistema operativo Android en su versión 4.2 o superior y en todas las escuelas donde existan mapeadores con la ayuda de un emulador de Android para PC y en coordinación con los técnicos de computación ejercitar a los niños en el conocimiento de los íconos de Mapa Verde. Éste juego de memoria visual fortalece y aumenta la capacidad de memorización de los mapeadores, de tal forma ayuda a mejorar la memoria en los niños de forma divertida.

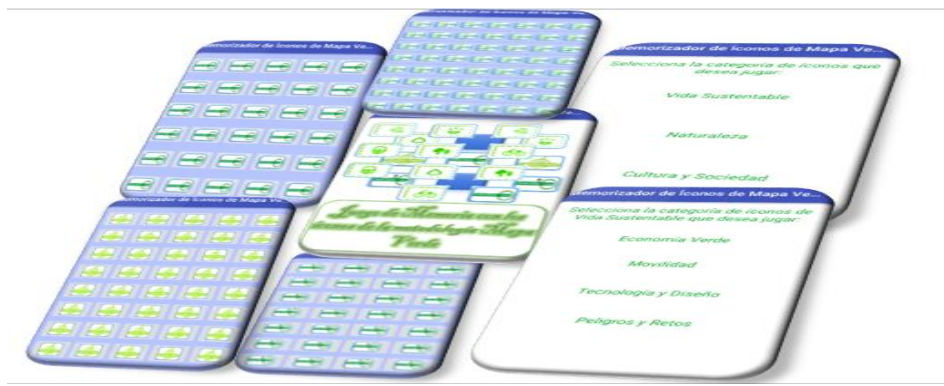

**Fig. 7.** Juego de memoria con los íconos de la metodología Mapa Verde. (Fuente: Elaboración propia)

En la figura 8 se muestran las capturas de pantalla del juego con los íconos de la metodología de Mapa Verde relacionadas con la explicación de algunas de las categorías de íconos de la metodología presentes en el juego.

**Tipografía Time New Roman 12 ptos.        Interlineado a 1.5      Justificación completa para el cuerpo**

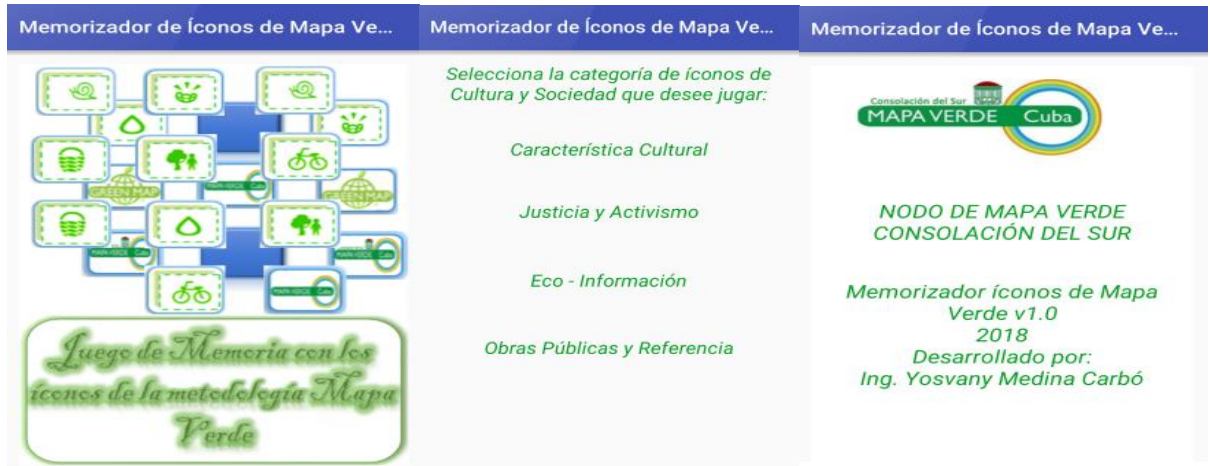

**Fig. 8.** Captura de pantalla del Juego de memoria con los íconos de la metodología Mapa Verde. (Fuente: Elaboración propia)

En la figura 9 se muestran las capturas de pantalla del juego con los íconos de la metodología de Mapa Verde relacionadas con la relación de íconos listos para jugar.

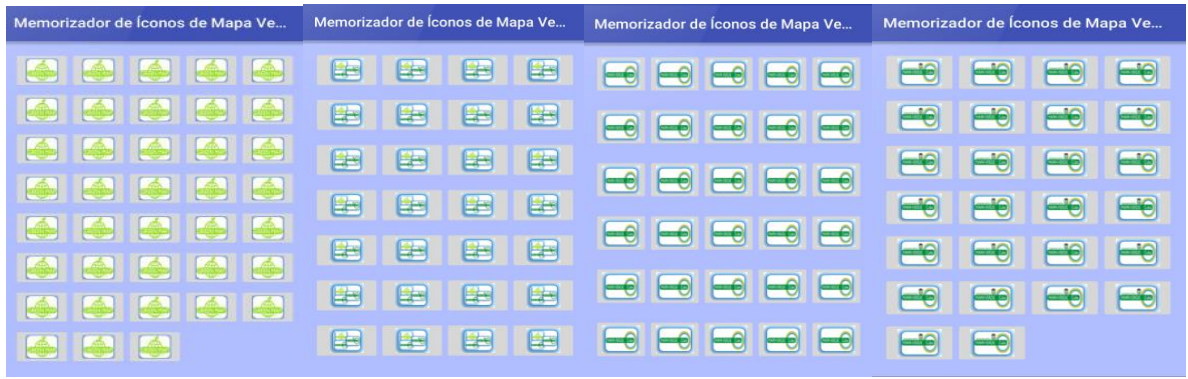

**Fig. 9.** Captura de pantalla de los niveles del juego. (Fuente: Elaboración propia)

En la figura 10 se muestran las capturas de pantalla del juego con los íconos de la metodología de Mapa Verde relacionadas con una imagen del juego en curso y una del juego terminado.

**Tipografía Time New Roman 12 ptos.        Interlineado a 1.5     Justificación completa para el cuerpo**

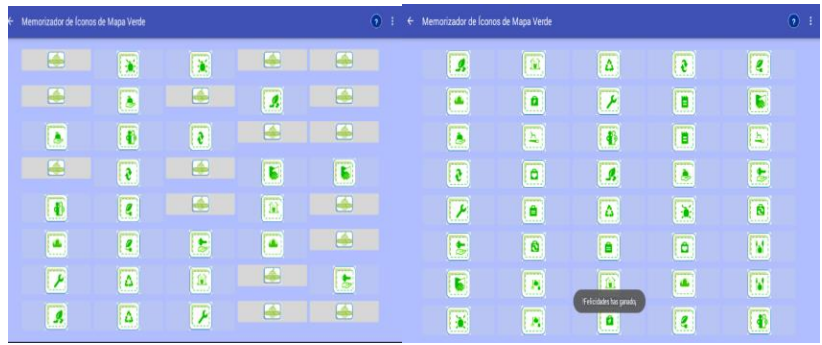

**Fig. 10.** Captura de pantalla del juego en curso y del juego terminado. (Fuente: Elaboración propia)

Con la utilización de estas herramientas tecnológicas y otros materiales entregados por el Centro Félix Varela se ha logrado insertar el tema relacionado con la metodología de Mapa Verde en todas las comunidades del municipio y con la ayuda de los actores sociales y de los decisores de cada comunidad se ha construido el Mapa Verde de cada consejo popular y de algunas de las comunidades del municipio. Además, se han formado mapeadores en las escuelas y creado círculos de interés y sociedades científicas en temas relacionados con la metodología y se han realizado sesiones de trabajo con los presidentes de los consejos populares donde se ha tomado como acuerdo la participación de los coordinadores en cada una de las asambleas con los delegados de circunscripción en cada consejo popular.

Como resultado de la utilización de estas herramientas tecnológicas desarrolladas en Android para dispositivos móviles, las sesiones de trabajo con los presidentes de los consejos populares, actores sociales, líderes comunitarios y la formación de mapeadores en las escuelas y en las comunidades se ha logrado desarrollar el Mapa Verde de los consejos populares y de algunas comunidades del municipio Consolación del Sur. Las ventajas que reporta es que sobre la base de la elaboración de los Mapas Verdes se logra la sensibilización de la comunidad y los actores sociales en temas relacionados con el cuidado y protección medio ambiente, además de ejecutar planes de acciones para lograr las posibles transformaciones en las comunidades, con la participación ciudadana en la solución de problemas ambientales, proyectos viables comunitarios con enfoque de género y equidad, procesos de concertación, incidencia en políticas

**Tipografía Time New Roman 12 ptos.  Interlineado a 1.5  Justificación completa para el cuerpo**

ambientales locales, relaciones de colaboración e intercambio y las alianzas con proyectos, instituciones y organismos.

# Conclusiones

Aunque a veces la tecnología pueda afectar negativamente al medio ambiente, bien utilizada, también puede servir para cuidar el planeta. Llevar un estilo de vida sostenible y saludable resulta cada vez más fácil y atractivo gracias a la creciente variedad de apps que surgen diariamente en el mercado. Resulta curioso, pero lo cierto es que la tecnología también puede ayudarte a cuidar el medio ambiente.

La realización del sistema de aplicaciones constituye un producto científico y una innovación tecnológica desarrollada para el estudio de la metodología de Mapa Verde y su posterior implementación en el municipio Consolación del Sur, la cual permitió:

- ➢ A los mapeadores apropiarse de herramientas y metodologías para el trabajo en las comunidades desde el Mapa Verde, como forma de garantizar la sostenibilidad.
- ➢ Sensibilizar a los pobladores de la comunidad en los principales problemas que se detectan, así como del cuidado y protección del entorno para preservar la salud. Para ello se desarrollan talleres de sensibilización, de capacitación a promotores y equipos de mapeadores.
- ➢ La utilización de una herramienta de educación ambiental que promueve el desarrollo de comunidades sustentables.

La utilización del mapa verde para promover una relación armónica entre las actividades del ser humano y su entorno, con la finalidad de garantizar la vida con calidad, mediante una estrategia que conlleva al establecimiento de alianzas a todos los niveles entre organismos e instituciones, llegándole al gobierno local, mediante la gestión de proyectos para el desarrollo local.

**Tipografía Time New Roman 12 ptos.** **Interlineado a 1.5** **Justificación completa para el cuerpo**

# Referencias

**Tipografía Time New Roman 12 ptos.	Interlineado a 1.5	Justificación completa para el cuerpo**

## Conflicto de interés

Ninguno de los autores manifestó la existencia de posibles conflictos de intereses que debieran ser declarados en relación con este artículo.

## Contribuciones de los autores

**Yosvany Medina Carbó:** contribución a la revisión bibliográfica, su análisis e interpretación. Desarrollo de la investigación. Redacción del borrador del artículo y de su versión final.

**Álvaro Celestino Alonso Vázquez:** contribución importante a la idea y diseño del estudio, revisión crítica del borrador del artículo y aprobación de la versión final a publicar.

**Reina María Rodríguez García:** contribución importante a la idea y diseño del estudio, revisión crítica del borrador del artículo y aprobación de la versión final a publicar.

## Financiación

**Tipografía Time New Roman 12 ptos.　　　Interlineado a 1.5　　Justificación completa para el cuerpo**

El trabajo no requirió financiación. El mismo forma parte del proyecto institucional "Implementación de la metodología de Mapa Verde para la gestión de la educación ambiental en comunidades consolareñas. " del Centro Universitario Municipal ¨Hermanos Saíz Monte de Oca¨ de Consolación del Sur.